\documentclass[preprint*]{JHEP3} 

\usepackage{epsfig,multicol,bbm}

\newcommand\fverb{\setbox\pippobox=\hbox\bgroup\verb}
\newcommand\fverbdo{\egroup\medskip\noindent%
            \fbox{\unhbox\pippobox}\ }
\newcommand\fverbit{\egroup\item[\fbox{\unhbox\pippobox}]}
\newbox\pippobox

\title{A Note on Noncommutative Brane Inflation}

\author{Xin Zhang\\
    Institute of Theoretical Physics, Chinese
Academy of Sciences, P.O.Box 2735, Beijing 100080, People's Republic
of China\\
    Interdisciplinary Center of Theoretical Studies, Chinese
Academy of Sciences, P.O.Box 2735, Beijing 100080, People's Republic
of China\\
    E-mail: \email{zhangxin@itp.ac.cn}}

\abstract{In this paper, we investigate the noncommutative KKLMMT
D3/$\overline{\rm D}$3 brane inflation scenario in detail.
Incorporation of the brane inflation scenario and the noncommutative
inflation scenario can nicely explain the large negative running of
the spectral index as indicated by WMAP three-year data and can
significantly release the fine-tuning for the parameter $\beta$.
Using the WMAP three year results (blue-tilted spectral index with
large negative running), we explore the parameter space and give the
constraints and predictions for the inflationary parameters and
cosmological observables in this scenario. We show that this
scenario predicts a quite large tensor/scalar ratio and what is
more, a too large cosmic string tension (assuming that the string
coupling $g_s$ is in its likely range from 0.1 to 1) to be
compatible with the present observational bound. A more detailed
analysis reveals that this model has some inconsistencies according
to the fit to WMAP three year results.}

\begin{document}


\section{Introduction}

The inflation paradigm provides a compelling explanation for the
homogeneity, isotropy and flatness of the universe by positing an
early epoch of accelerated expansion \cite{Guth}. This accelerated
period of expansion also predicts an almost scale-invariant density
perturbation power spectrum, which has received strong observational
support from the measurement of the temperature fluctuation in the
cosmic microwave background radiation \cite{cobe,wmap1}. It is
expected for inflation that it should emerge naturally from some
fundamental theory of microphysics. String theory is the most
promising candidate hitherto for a fundamental theory, so it is
rather natural to look for an explicit realization of the
inflationary scenario within the framework of the superstring
theory. An attractive scenario for inflation based upon string
theory is the brane-antibrane inflation
\cite{Dvali:1998pa,Burgess:2001fx} which proposes that inflation
might arise from the interaction potential between a D3-brane and an
anti-D3-brane which are parallel and widely separated in a six
dimensional compact manifold where the six compactified dimensions
are dynamical stabilized \cite{Giddings:2001yu,kklt}. The
anti-D3-brane is held fixed at the bottom of a warped throat (due to
an attractive force) while the D3-brane is mobile. The D3-brane
experiences a small attractive force towards the anti-D3-brane. The
distance between the branes plays the role of the inflaton field.
This is the KKLMMT scenario \cite{kklmmt}. Herein, inflation ends
when the D3-brane and the anti-D3-brane collide and annihilate,
initiating the hot big bang epoch. The annihilation of the
D3-$\overline{\rm D}$3-branes allows the universe to settle down to
the string vacuum state that describes our universe. During the
process of the brane collision, cosmic strings are copiously
produced \cite{Jones:2002cv,Sarangi:2002yt}. For a well-worked
inflationary scenario, the production of topological defects other
than cosmic strings must be suppressed by many orders of magnitude.
It should be notable that the D3/$\overline{\rm D}$3 brane inflation
scenario achieves this property automatically.

\begin{figure}[htbp]
\begin{center}
\includegraphics[scale=1.2]{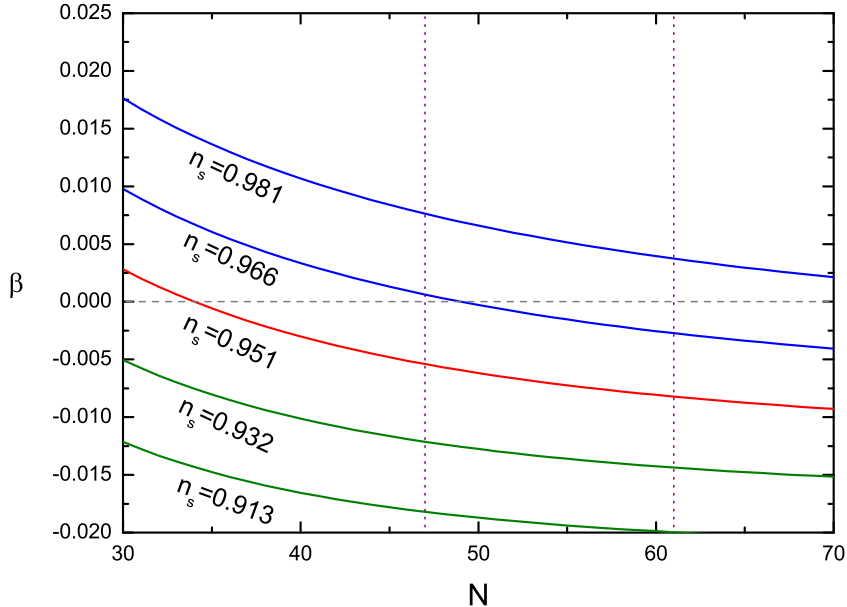}
\caption[]{\small The constraint relationship between the parameter
$\beta$ and the $e$-folding number $N$ before the end of inflation,
in the usual (i.e. commutative) D3-${\rm \overline{D}3}$ brane
inflation, according to the WMAP results of the spectral index of
the primordial density perturbation (within 2 standard deviations)
which comes from assuming a power law spectrum. Note that the area
of $\beta>0$ and $47<N<61$ is a rational region for the brane
inflation.}\label{fig:nbpl}
\end{center}
\end{figure}

For the D3/$\overline{\rm D}$3 brane inflation scenario, the
potential that arises from the attraction between the brane and the
antibrane in the throat generically takes the following form
\cite{Firouzjahi:2005dh}
\begin{equation}
V(\phi)={1\over 2}\beta H^2\phi^2+2T_3h^4\left
(1-{M^4\over\phi^4}\right), \label{potential}
\end{equation}
where
\begin{equation}
M^4={27\over 32\pi^2}T_3h^4.\label{m4}
\end{equation}
Here, $T_3$ is the tension of the D3-brane, $h$ is the warp factor
of the throat, $H$ is the Hubble parameter, and $\phi$ is the
canonical inflaton ($\phi=\sqrt{T_3}r$, where $r$ is the position of
the D3-brane with respect to the bottom of the throat). In equation
(\ref{potential}), the first term ($\beta$ term, where $\beta$
characterizes the mass of the inflaton field) comes from the
contributions of the K\"ahler potential and various interactions in
the superpotential \cite{kklmmt} as well as possible D-terms
\cite{Burgess:2003ic}, the second term ($T_3h^4$ term) is the
effective cosmological constant coming from the presence of the
$\overline{\rm D}$3-brane (with tension $T_3$) sitting at the bottom
of the throat, and the last term ($\phi^{-4}$ term) is the
Coulombic-like attractive potential between the D3-brane and the
$\overline{\rm D}$3-brane \cite{kklmmt}. The parameter $\beta$ is
generically of order unity. However, to achieve slow roll, $\beta$
has to be fine-tuned to be much smaller than 1. This case
corresponds to the KKLMMT scenario, where the effect of the $\beta$
term is negligible. Effort has been made for relaxing the
fine-tuning of $\beta$ \cite{Firouzjahi:2005dh}. As is shown in
reference \cite{Firouzjahi:2005dh} that the fine-tuning of $\beta$
can be relaxed to $\beta\lesssim 1/7$ when admitting $n_s\leq 1.086$
(2 $\sigma$ result) which comes from an analysis of the first year
WMAP data combined with SDSS Lyman $\alpha$ and other data
\cite{Seljak:2004xh}. What is remarkable is that slow roll inflation
still works fine in this case, while the cosmic string tension
increases rapidly. In this very simple generalization of the KKLMMT
scenario, the cosmic string tension can easily saturate the
observational bound. As the fine-tuning is relaxed, the primordial
density perturbation spectrum moves from red-tilted to blue-tilted.
However, the recent announced WMAP three-year data explicitly
indicate that the scalar spectral index of the primordial
fluctuations is red-tilted when assuming a power law spectrum,
namely $n_s=0.951^{+0.015}_{-0.019}$ \cite{wmap3}. This fact places
significant and stringent constraints on the D3/$\overline{\rm D}$3
brane inflation scenario \cite{Huang:2006ra}. Here we show in figure
\ref{fig:nbpl} the constraints on the brane inflation model from the
WMAP result of the spectral index of the primordial density
perturbation. Note that $\beta<0$ is forbidden since in this case
$\beta$ term in the potential tends to push D3-brane out of the
throat such that the inflation will not happen. However, we see
clearly in this figure that according to the WMAP result $\beta$ is
very likely to be less than zero. For example, $n_s=0.951$ gives
$\beta=-0.007$ at $N=54$, where $N$ is the $e$-folds before the end
of the inflation. It should be pointed out that a reasonable range
for $e$-folds is $47<N<61$ (or $N=54\pm 7$) \cite{Alabidi:2005qi}.
Thus the area of $\beta>0$ and $47<N<61$ seems a rational region for
the brane inflation. The curve for $n_s=0.966$ (1 $\sigma$) has a
very small part locating in this region as shown in figure
\ref{fig:nbpl}. For $n_s=0.966$, taking $N=47$ one derives
$\beta=6.2\times 10^{-4}$, this is a fine-tuning. The curve for 2
$\sigma$ result $n_s=0.981$ (note that we take 2 $\sigma$ as twice
of 1 $\sigma$ here for simplicity) resides in the legal region. Here
for this case we take $N=54$ and then get $\beta=5.4\times 10^{-3}$
which is also a fine-tuning value. On the whole, even though
$\beta>0$ can be achieved constrainedly, the fine-tuning of $\beta$
is inevitable provided that the power spectrum of the primordial
perturbation is red-tilted.

On the other hand, for the power spectrum of primordial fluctuations
there exists the possibility that the spectral index varies slowly
with scale. When allowing for a running spectral index, the WMAP
three-year data favor a blue-tilted spectral index with a large
negative running (and also a large tensor amplitude, characterized
by $r$, the ratio of the tensor to scalar power spectrum)
\cite{wmap3},
\begin{equation}
n_s=1.21^{+0.13}_{-0.16},~~~~\alpha_s=dn_s/d\ln
k=-0.102^{+0.050}_{-0.043},~~~~r\leqslant 1.5~~~{\rm
at~95\%~CL}.\label{wmap3}
\end{equation}
However, the D3-$\overline{\rm D}$3 brane inflation scenario can not
produce a so large absolute value of the running spectral index (the
typical value it attains is $|\alpha_s|\sim 10^{-4}$) though the
blue-tilted spectrum can release the fine-tuning of $\beta$
effectively. Actually, it is rather hard for most inflationary
models to produce such a large running of the spectral index within
the framework of the usual slow-roll mechanism \cite{hardrun}.
However, it has been shown that when considering the space-time
noncommutative effects in the very early universe due to the
possible effects of quantum gravity, the inflation models can nicely
produce a large running spectral index
\cite{noncomu,ncnew}.\footnote{For extensive studies on the
noncommutative inflation models see \cite{ncold}, and for other
discussions on the running spectral index see \cite{run1,run2}.}
Recently, a noncommutative brane inflation scenario has been
proposed to accommodate the large negative running of the spectral
index in the WMAP three-year results and to release the fine-tuning
of $\beta$ due to the blue-tilted spectrum \cite{Huang:2006zu}. In
this paper, we will investigate this noncommutative brane inflation
model in detail. We shall show the ranges of the mass parameter
$\beta$ as well as the noncommutative parameter $\mu$ allowed by
WMAP three-year results of the spectral index and its running.
Within the allowed ranges of $\beta$ and $\mu$, we evaluate the
tensor-to-scalar ratio and the cosmic string tension. We shall show
that such a noncommutative brane inflation will predict a rather
large tensor-to-scalar ratio (implying a large gravity wave
amplitude) and what is more, a too large cosmic string tension
(assuming the value of the string coupling $g_s$ is in the range 0.1
to 1) to be compatible with the present observational results of the
anisotropy of the cosmic microwave background radiation.
Furthermore, we analyze the various parameters and physical
quantities/cosmological observables in this noncommutative brane
inflation model by fitting to the three-year WMAP results and show
that some inconsistencies occur in this model. We shall discuss
about these inconsistencies and give some comments.

\section{A model of noncommutative brane inflation}

The space-time noncommutative effects can arise naturally from the
string theory \cite{Seiberg:2000ms,Gopakumar:2000na}. The space and
time coordinates will not commute with each other at very high
energies in the early universe when quantum effects are expected to
be important. This noncommutativity leads to an uncertainty relation
between time and space, which was advocated as a generic property of
string theory \cite{uncty},
\begin{equation}
\Delta t\Delta x_{\rm phys}\geqslant L_{\rm nc}^2,
\end{equation}
where $t$, $x_{\rm phys}$ are the physical space-time coordinates
and $L_{\rm nc}$ is the characteristic scale of noncommutative
effects. The space-time noncommutative effects can be encoded in a
new product, star product, replacing the usual algebra product
\cite{Seiberg:2000ms,Gopakumar:2000na}. The evolution of a
homogeneous and isotropic background will not change and the
standard cosmological equations based upon the
Friedmann-Robertson-Walker (FRW) metric remain the same:
\begin{equation}
\ddot{\phi}+3H\dot{\phi}+V'(\phi)=0,
\end{equation}
\begin{equation}
H^2=\left(\dot{a}\over a\right)^2={1\over 3M_{\rm
Pl}^2}\left({1\over2}\dot{\phi}^2+V(\phi)\right),
\end{equation}
where $M_{\rm Pl}$ is the reduced Planck mass. If $\dot{\phi}^2\ll
V(\phi)$ and $\ddot{\phi}\ll 3H\dot{\phi}$, the scalar field shall
slowly roll down its potential. Define some slow-roll parameters,
\begin{equation}
\epsilon=-{\dot{H}\over H^2}={M_{\rm Pl}^2\over 2}\left(V'\over
V\right)^2,
\end{equation}
\begin{equation}
\eta=\epsilon-{\ddot{H}\over 2H\dot{H}}=M_{\rm Pl}^2{V''\over V},
\end{equation}
\begin{equation}
\xi^2=7\epsilon\eta-5\epsilon^2-2\eta^2+\zeta^2=M_{\rm
Pl}^4{V'V'''\over V^2},
\end{equation}
where $\zeta^2=\stackrel{...}{H}/(2H^2\dot{H})$, then the slow-roll
condition can be expressed as $\epsilon, |\eta|\ll 1$. The inflation
ends when the slow-roll condition ceases to be satisfied.

The stringy space-time uncertainty relation leads to changes in the
action for the metric fluctuations \cite{Brandenberger:2002nq}. The
action for scalar metric fluctuations can be reduced to the action
of a real scalar field $u$ with a specific time-dependent mass which
depends on the background cosmology. Thus the modified action of the
perturbation led by the stringy space-time uncertainty relation can
be expressed as \cite{Brandenberger:2002nq}
\begin{equation}
S=V_T\int\limits_{k<k_0}d\tilde{\eta}d^3k{1\over
2}z_k^2(\tilde{\eta})(u'_{-k}u'_k-k^2 u_{-k}u_k),
\end{equation}
where $V_T$ is the total spatial coordinate volume, a prime denotes
the derivative with respect to the modified conformal time
$\tilde{\eta}$, $k$ is the comoving wave number, and
\begin{equation}
z_k^2(\tilde{\eta})=z^2(\beta_k^+\beta_k^-)^{1/2},~~~~z={a\dot{\phi}\over
H},
\end{equation}
\begin{equation}
{d\tilde{\eta}\over
d\tau}=\left(\beta_k^-\over\beta_k^+\right)^{1/2},~~~~\beta_k^\pm={1\over
2}(a^{\pm 2}(\tau+kL_{\rm nc}^2)+a^{\pm 2}(\tau-kL_{\rm nc}^2)),
\end{equation}
where $k_0=(\beta_k^+/\beta_k^-)^{1/4}L_{\rm nc}$, and $\tau$
denotes a new time variable in terms of which the stringy
uncertainty principle takes the simple form $\Delta\tau\Delta
x\geqslant L_{\rm nc}^2$, using comoving coordinates $x$. The case
of commutative space-time corresponds to $L_{\rm nc}=0$. The
comoving curvature perturbation is defined as ${\cal R}\equiv
\Phi+H\delta\phi/\dot{\phi}$, where $\Phi$ is the scalar metric
perturbation in longitudinal gauge, and then we have ${\cal
R}_k(\tilde{\eta})=u_k(\tilde{\eta})/z_k(\tilde{\eta})$. The
equation of motion for the scalar perturbation can be written as
\begin{equation}
u''_k+\left(k^2-{z''_k\over z_k}\right)u_k=0.\label{eomcurv}
\end{equation}
After a lengthy but straightforward calculation, we get
\begin{equation}
{z''_k\over z_k}=2(aH)^2\left(1+{5\over 2}\epsilon-{3\over
2}\eta-2\mu\right),
\end{equation}
\begin{equation}
aH\simeq{-1\over\tilde{\eta}}(1+\epsilon+\mu),
\end{equation}
where $\mu=H^2k^2/(a^2M_{\rm nc}^4)$ is the noncommutative
parameter, and $M_{\rm nc}=L_{\rm nc}^{-1}$ is the noncommutative
mass scale.

The power spectrum on superhorizon scales of the comoving curvature
perturbation in the noncommutative space-time can be obtained by
solving the equation of motion (\ref{eomcurv}) \cite{noncomu},
\begin{equation}\label{ncps}
\Delta_{\cal R}^2\simeq {k^3\over 2\pi^2}|{\cal
R}_k(\tilde{\eta})|^2\simeq {1\over 2\epsilon}{1\over M_{\rm
Pl}^2}\left(H\over 2\pi\right)^2(1+\mu)^{-4-6\epsilon+2\eta},
\end{equation}
here $H$ takes the value when the fluctuation mode $k$ crosses the
Hubble radius ($z''_k/z_k=k^2$). Then the spectral index of the
power spectrum of the scalar fluctuations and its running can be
given,
\begin{equation}\label{index}
n_s-1\equiv {d\ln \Delta_{\cal R}^2\over d\ln
k}=-6\epsilon+2\eta+16\epsilon\mu,
\end{equation}
\begin{equation}\label{running}
\alpha_s\equiv{d n_s\over d\ln k}
=-24\epsilon^2+16\epsilon\eta-2\xi^2-32\epsilon\eta\mu.
\end{equation}
When $L_{\rm nc}\rightarrow 0$ or $M_{\rm nc}\rightarrow +\infty$,
the noncommutative parameter $\mu=H^2k^2/(a^2M_{\rm
nc}^4)\rightarrow 0$, Eqs. (\ref{index}) and (\ref{running})
reproduce the results of the commutative case. The tensor-to-scalar
ratio is also given,
\begin{equation}\label{tensor}
r\equiv{\Delta_{\rm grav}^2\over\Delta_{\cal R}^2}=16\epsilon.
\end{equation}

Following the proposal of reference \cite{Huang:2006zu}, one can
embed the D3/$\overline{D}$3 brane inflation into the noncommutative
inflation framework. In the brane inflation scenario, the slow roll
parameter $\eta$ can be evaluated \cite{Firouzjahi:2005dh}
\begin{equation}\label{eta}
\eta={\beta\over 3}-20M_{\rm Pl}^2{M^4\over \phi^6}.
\end{equation}
It should be pointed out that in the derivation of the $\eta$
expression (\ref{eta}) one has used the fact that parameter $\beta$
is small enough such that the approximation $V\sim 2T_3h^4$ can be
achieved. The end of inflation is determined by the condition that
the slow roll parameters cease to be small. Thus one can use the
condition $\eta\sim -1$ to terminate the inflation, which gives the
final value of inflaton field
\begin{equation}\label{phiend}
\phi_{\rm end}^6={20 M_{\rm Pl}^2 M^4\over 1+\beta/3}.
\end{equation}
The number of $e$-folds before the end of inflation is given by
\begin{equation}
N={1\over M_{\rm Pl}^2}\int\limits_{\phi_{\rm end}}^{\phi_N} {V\over
V'}d\phi,
\end{equation}
where $\phi_N$ denotes the value of $\phi$ at the time of
horizon-crossing when the $e$-folding number equals to $N$, hence we
have
\begin{equation}\label{npp}
e^{2\beta N}={\beta\phi_N^6+12M_{\rm Pl}^2M^4\over\beta\phi_{\rm
end}^6+12M_{\rm Pl}^2M^4}.
\end{equation}
Note that to get this formula the parameter $\beta$ is required to
be small enough. Then combining (\ref{phiend}) and (\ref{npp}) we
can derive
\begin{equation}
\phi_N^6=24NM_{\rm Pl}^2M^4\Omega(\beta),\label{phin}
\end{equation}
where
\begin{equation}
\Omega(\beta)={(1+2\beta)e^{2\beta N}-(1+\beta/3)\over
2\beta(N+5/6)(1+\beta/3)}.
\end{equation}

\begin{figure}[htbp]
\centering
\begin{center}
$\begin{array}{c@{\hspace{0.2in}}c} \multicolumn{1}{l}{\mbox{}} &
\multicolumn{1}{l}{\mbox{}} \\
\includegraphics[scale=0.68]{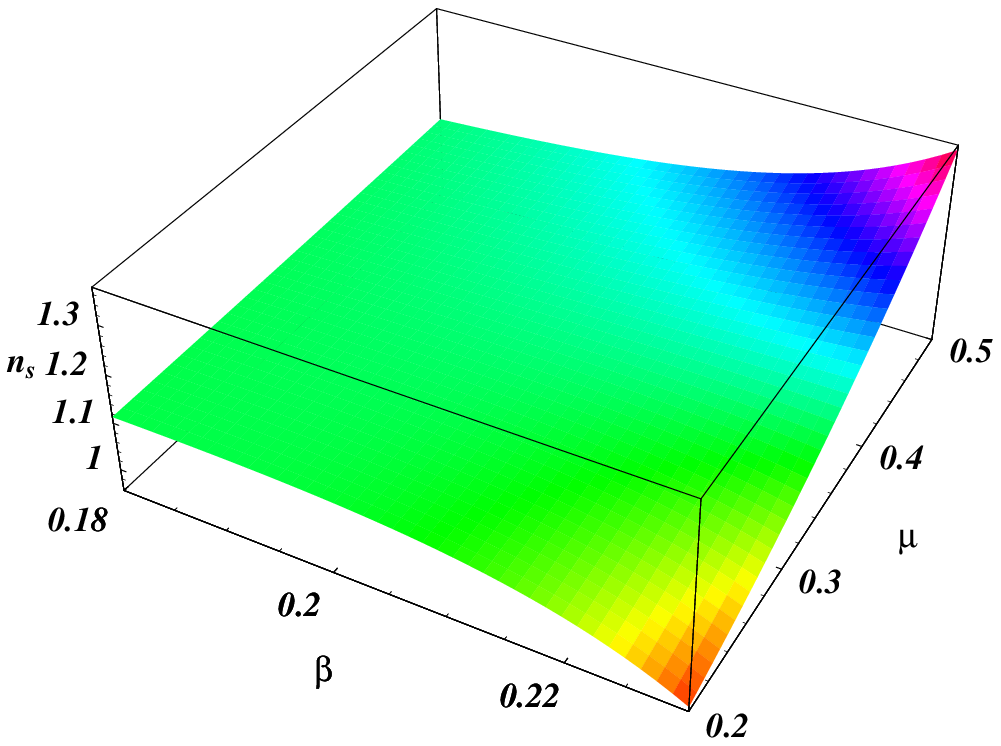} &\includegraphics[scale=0.68]{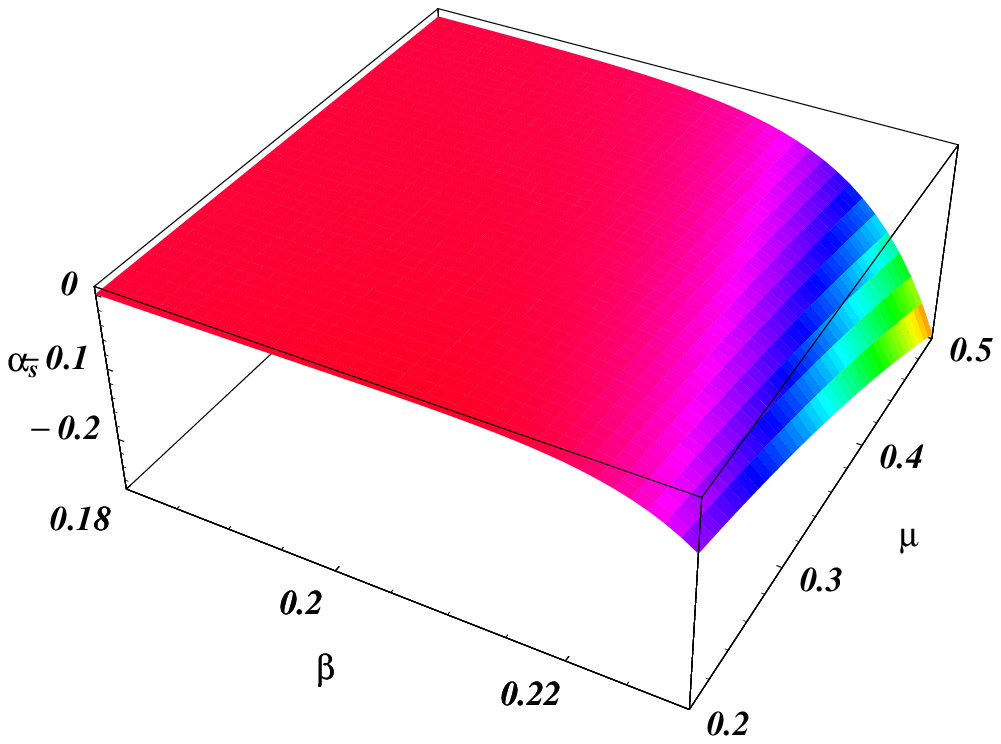} \\
\mbox{\bf (a)} & \mbox{\bf (b)}
\end{array}$
$\begin{array}{c@{\hspace{0.2in}}c} \multicolumn{1}{l}{\mbox{}} &
\multicolumn{1}{l}{\mbox{}} \\
\includegraphics[scale=0.68]{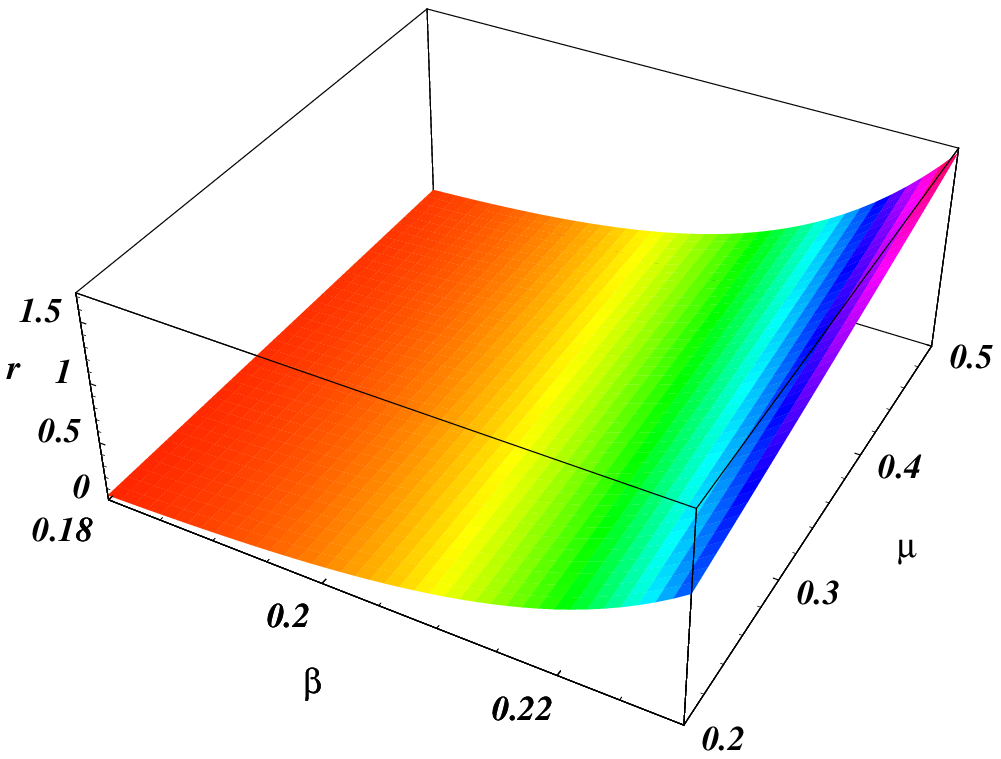} &\includegraphics[scale=0.68]{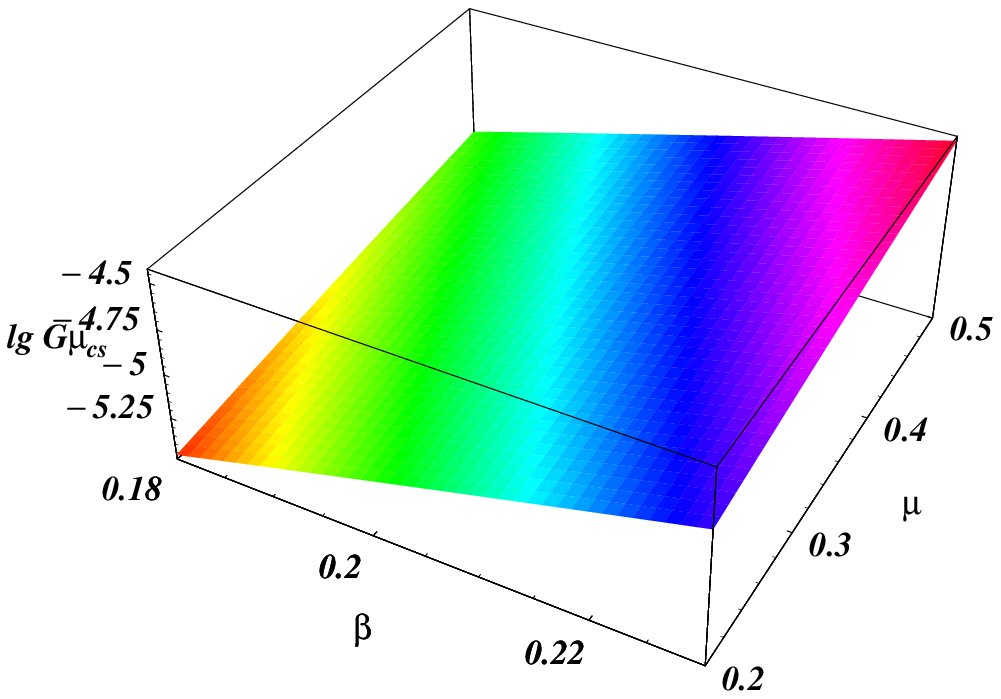} \\
\mbox{\bf (c)} & \mbox{\bf (d)}
\end{array}$
\end{center}
\caption[]{\small The spectral index, its running, the
tensor-to-scalar ratio and the cosmic string tension as functions of
$\beta$ and $\mu$, when the $e$-folds before the end of inflation
$N=54$, in the noncommutative D3/$\overline{\rm D}$3 brane inflation
scenario.} \label{fig:quants}
\end{figure}

\begin{figure}[htbp]
\centering
\begin{center}
$\begin{array}{c@{\hspace{0.2in}}c} \multicolumn{1}{l}{\mbox{}} &
\multicolumn{1}{l}{\mbox{}} \\
\includegraphics[scale=0.8]{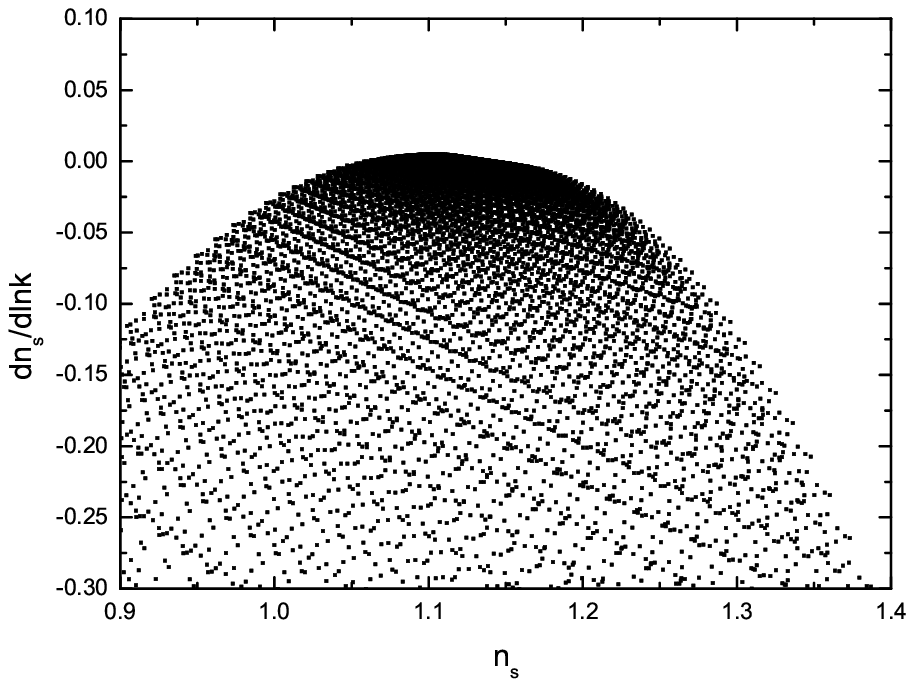} &\includegraphics[scale=0.8]{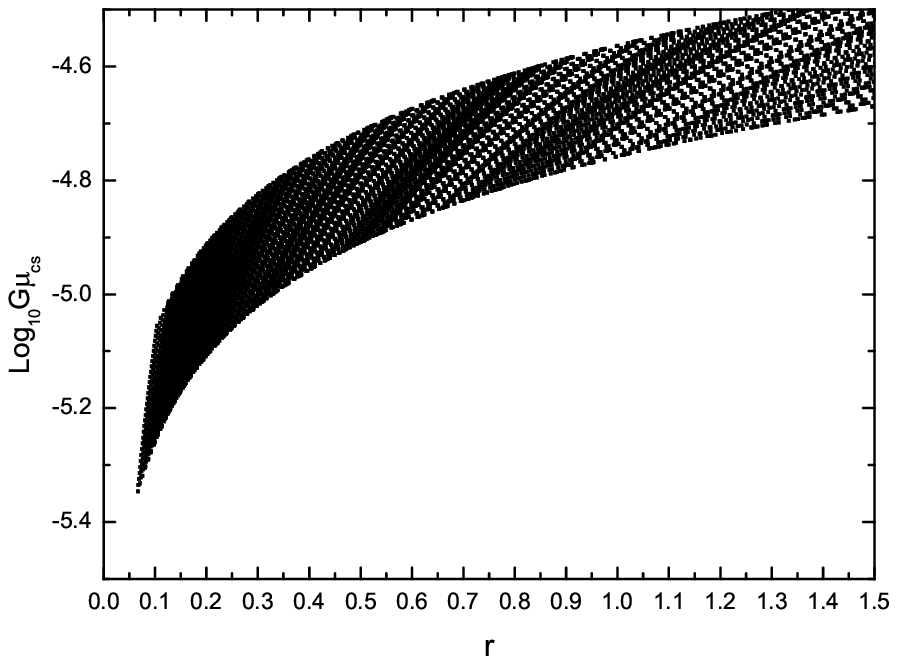} \\
\mbox{\bf (a)} & \mbox{\bf (b)}
\end{array}$
\end{center}
\caption[]{\small Correlations between observables in the model. The
points come from varying the parameters $N\in (50, 60)$, $\beta\in
(0.20, 0.25)$ and $\mu\in (0.2, 0.5)$. Note that we only show the
appropriate part of the whole region. Panel (a): The running of the
spectral index $dn_s/d\ln k$ versus the spectral index $n_s$. Panel
(b): The logarithm of the cosmic string tension $\lg (G\mu_{cs})$
versus the tensor-to-scalar ratio $r$.} \label{fig:correlation}
\end{figure}

When considering the space-time noncommutative effects, the power
spectrum of scalar curvature perturbations $\cal R$ (i.e. the power
spectrum of primordial density perturbations) is given by
\cite{Huang:2006zu}
\begin{equation}
\Delta_{\cal R}^2\simeq {V/M^4_{\rm Pl}\over
24\pi^2\epsilon}(1+\mu)^{-4}=\left({2^5\over
3\pi^4}\right)^{1/3}\left({T_3h^4\over M^4_{\rm
Pl}}\right)^{2/3}N^{5/3}f(\beta)^{-4/3}(1+\mu)^{-4},\label{power}
\end{equation}
where
\begin{equation}
f(\beta)=\Omega(\beta)^{-5/4}(1+2\beta N\Omega(\beta))^{3/2},
\end{equation}
here physical quantities all take the values of Hubble radius
crossing (Hubble exit during inflation). Note that the COBE
normalization \cite{cobenorm} corresponds to $\Delta_{\cal
R}^2\simeq 2\times 10^{-9}$ for the mode which crosses the Hubble
radius about 55 $e$-folds before the end of inflation. Comparing
(\ref{power}) with (\ref{ncps}), we find a reasonable approximation
has been made due to that $\mu$, $\epsilon$ and $\eta$ are all small
quantities and $-6\epsilon+2\eta$ appears in the power. So, it can
be seen clearly that the space-time noncommutative effects enter the
primordial power spectrum (\ref{power}) through the factor
$(1+\mu)^{-4}$. The original KKLMMT scenario can be reproduced when
$\mu\rightarrow 0$ and $\beta\rightarrow 0$. We shall see below that
the noncommutative effects encoded in the parameter $\mu$ also enter
the slow-roll parameter $\epsilon$ so as to affect the observational
quantities of interest. Combining equations (\ref{m4}), (\ref{phin})
and (\ref{power}), we obtain
\begin{equation}
{T_3h^4\over M_{Pl}^4}=\left({3\pi^4\over
2^5}\right)^{1/2}(\Delta_{\cal
R}^2)^{3/2}N^{-5/2}f(\beta)^2(1+\mu)^6,
\end{equation}
\begin{equation}\label{phine}
{\phi_N\over M_{\rm Pl}}=\left({27\over 8}\right)^{1/4}(\Delta_{\cal
R}^2)^{1/4}N^{-1/4}f(\beta)^{1/3}\Omega(\beta)^{1/6}(1+\mu).
\end{equation}
Now the slow-roll parameters can be written as
\begin{equation}
\epsilon={1\over 18}\left({\phi_N\over M_{\rm
Pl}}\right)^2\left(\beta+{1\over 2N\Omega(\beta)}\right)^2,
\end{equation}
\begin{equation}
\eta={\beta\over 3}-{5\over 6}{1\over N\Omega(\beta)},
\end{equation}
\begin{equation}
\xi^2={5\over 3N\Omega(\beta)}\left(\beta+{1\over
2N\Omega(\beta)}\right).
\end{equation}
Using equation (\ref{phine}), the slow-roll parameter $\epsilon$ can
be explicitly expressed in terms of $\mu$,
\begin{equation}
\epsilon={1\over 4\sqrt{6N}}(\Delta_{\cal
R}^2)^{1/2}\Omega(\beta)^{1/3}f(\beta)^{2/3}\left(\beta+{1\over
2N\Omega(\beta)}\right)^2(1+\mu)^2.
\end{equation}
Hence, in the noncommutative D3/$\overline{D}$3 brane inflation
scenario, the observational quantities of interest, such as the
spectral index $n_s$, its running $\alpha_s$ and the
tensor-to-scalar ratio $r$ given by equations (\ref{index}),
(\ref{running}) and (\ref{tensor}), can all be expressed as
functions of parameters $N$, $\beta$ and $\mu$, given COBE
normalization $\Delta_{\cal R}^2=2\times 10^{-9}$. When fixing the
$e$-folding number before the end of inflation $N$ at some
reasonable value, we can show explicitly the predictions of the
above inflationary quantities within some ranges of $\beta$ and
$\mu$. For example, fixing $N=54$, the functions $n_s(\beta,\mu)$,
$\alpha_s(\beta,\mu)$ and $r(\beta,\mu)$ are plotted in figure
\ref{fig:quants}, shown as (a), (b) and (c), respectively. Due to
the contributions of the noncommutative parameter $\mu$, the large
negative running of the spectral index can be successfully achieved
and moreover, the fine-tuning of $\beta$ in the KKLMMT model can be
released here roughly to $1/5-1/4$. It is clearly that a blue-tilted
spectral index with a large negative running can be realized nicely
in the noncommutative D3/$\overline{D}$3 brane inflation model, when
$\beta$ and $\mu$ locate at some appropriate values such as
$\beta\sim 0.225$ and $\mu\sim 0.43$. Meanwhile, figure
\ref{fig:quants} explicitly shows that this model can produce a
rather large gravity wave $r\sim 1.1$. In addition, we show in
figure \ref{fig:correlation} the correlations between observables in
the model. The points come from varying the parameters $N\in (50,
60)$, $\beta\in (0.20, 0.25)$ and $\mu\in (0.2, 0.5)$. The panel (a)
in figure \ref{fig:correlation} shows the correlation between $n_s$
and $\alpha_s$, and we see that the WMAP three-year results,
$n_s=1.21^{+0.13}_{-0.16}$ and $\alpha_s=-0.102^{+0.050}_{-0.043}$,
are covered by the model.

\section{Cosmic string tension}

Cosmic strings and other topological defects have long been proposed
as one of the candidates for the origin of structure formation
\cite{Zeldovich:1980gh}, however, it has been shown that this
scenario leads to predictions incompatible with observations such as
the power spectrum of cosmic microwave background (CMB) temperature
anisotropy \cite{Pen:1997ae}. Nevertheless, the brane/anti-brane
inflation scenario inspired from string theory naturally indicates
that cosmic strings would have a small contribution to the CMB,
which is compatible with the observational limits
\cite{Jones:2002cv,Sarangi:2002yt}. This has led to a significant
revival of all aspects of cosmic string scenario, including new
theoretical motivations, phenomenological implications and direct
observational searches, see e.g.
\cite{Firouzjahi:2005dh,Pogosian:2003mz,Copeland:2003bj,Leblond:2004uc,Jackson:2004zg,
Damour:2004kw,Shandera:2006ax,Seljak:2006hi,Wyman:2005tu}. The
possibility of detecting the signal of the cosmic strings through
astronomical observations opens a significant window to test string
theory. The current observational bound (the upper limit) for cosmic
string tension from the CMB temperature is roughly $G\mu_{\rm
obs}\lesssim 2.7\times 10^{-7}$ at $95\%$ confidence level
($1.8\times 10^{-7}$ at $68\%$ confidence level)
\cite{Wyman:2005tu}. In the original KKLMMT scenario (corresponds to
$\mu\rightarrow 0$, $\beta\rightarrow 0$), the tension of these
cosmic strings is naturally predicted to be small and their
contribution to CMB is well below the current limits
\cite{Firouzjahi:2005dh,Copeland:2003bj,Shandera:2006ax}. Within the
reasonable narrow range of parameter space after WMAP3, for example,
taking $N=47$ and $\beta=6.2\times 10^{-4}$ (this case corresponds
to $n_s=0.966$), one derives the corresponding cosmic string tension
$G\mu_{\rm th}=4.43\times 10^{-10}$ (assuming the string coupling
$g_s=1$). In this section, we shall investigate the cosmic string
tension in the noncommutative brane inflation model.

At the end of inflation, the D3-brane collides with the
anti-D3-brane at the bottom of the throat. The annihilation of the
D3-$\overline{\rm D}3$-branes initiates the hot big bang epoch,
meanwhile D1-branes (i.e. D-strings) and fundamental closed strings
(i.e. F-strings) are also produced. The quantities of interest are
the D-string tension $G\mu_D$ and the F-string tension $G\mu_F$,
where $\mu_D$ and $\mu_F$ are the effective tensions measured from
the viewpoint of the four dimensional effective action. Since in ten
dimensions, there is the relationship $T_3=T_F^2/2\pi
g_s=T_D^2g_s/2\pi$, where $T_3=1/(2\pi)^3g_s\alpha'^2$ is the
D3-brane tension, we have for the tensions of the IIB string in the
inflationary throat
\begin{equation}
G\mu_F=GT_Fh^2=\sqrt{g_s\over 32\pi}\left({T_3h^4\over M_{\rm
Pl}^4}\right)^{1/2},
\end{equation}
\begin{equation}
G\mu_D=GT_Dh^2=\sqrt{1\over 32\pi g_s}\left({T_3h^4\over M_{\rm
Pl}^4}\right)^{1/2},
\end{equation}
where $G=1/(8\pi M_{\rm Pl}^2)$ is the four dimensional Newton's
gravity constant and $g_s$ is the string coupling. There are also
bound states of $p$ F-strings and $q$ D-strings with a cosmic string
network spectrum \cite{Copeland:2003bj}
\begin{equation}
\mu_{(p,q)}=\mu_F\sqrt{p^2+q^2/g_s^2}.
\end{equation}
We see that both F-strings and D-strings are dependent on the string
coupling $g_s$, but the geometric mean $(\mu_F\mu_D)^{1/2}$ is
independent of $g_s$. Thus, for convenience, we define
\begin{equation}
G\mu_{cs}=G(\mu_F\mu_D)^{1/2}=\sqrt{1\over 32\pi}\left({T_3h^4\over
M_{\rm Pl}^4}\right)^{1/2}.
\end{equation}
Then, combining with equation (\ref{power}), one obtains the cosmic
string tension in the noncommutative D3/$\overline{D}$3-brane
inflation model
\begin{equation}\label{cst}
G\mu_{cs}=\left({3\pi^2\over 2^{15}}\right)^{1/4}(\Delta_{\cal
R}^2)^{3/4}N^{-5/4}f(\beta)(1+\mu)^3.
\end{equation}
Obviously, for F-string we have $\mu_F=\mu_{cs}\sqrt{g_s}$, and for
D-string we have $\mu_D=\mu_{cs}/\sqrt{g_s}$.

Next, we show in the panel (d) of figure \ref{fig:quants} the cosmic
string tension $G\mu_{cs}$ (logarithm) of the noncommutative
D3/$\overline{D}$3-brane inflation scenario as function of $\beta$
and $\mu$, where the number of $e$-folds before the end of inflation
is fixed at $N=54$. We see clearly in figure \ref{fig:quants} that
the tension of cosmic strings produced in the noncommutative brane
inflation is unbelievable large. When parameters $\beta$ and $\mu$
are taken as some appropriate values, for instance $\beta\sim 0.225$
and $\mu\sim 0.43$, the model can realize a reasonable result for
the power spectrum of the primordial perturbations, namely a
blue-tilted spectral index with a large negative running, as
indicated by WMAP three-year data \cite{wmap3}. However, as the
primordial power spectrum is realized successfully, the cosmic
strings produced by the model seem not reasonable. Taking $N=54$,
$\beta=0.225$ and $\mu=0.43$, we obtain $\lg(G\mu_{cs})=-4.58$.
Recall that the observational upper limit on the cosmic string
tension is $G\mu_{\rm obs}<2.7\times 10^{-7}$ (at 95$\%$ confidence
level), namely $\lg (G\mu_{\rm obs})<-6.75$, so the predicted
tension of cosmic strings (assuming $g_s=1$, the F-strings and
D-strings are in coincidence) is larger than the observational bound
by roughly two orders of magnitude. Assuming the string coupling
$g_s=0.1$ (note that this is a representative value for
$g_s$)\footnote{The value of $g_s$ is likely in the range 0.1 to 1
\cite{Copeland:2003bj}. Note that $g_s>1$ can be converted to
$g_s<1$ by $S$-duality.} as adopted usually in the
literature,\footnote{See e.g. references
\cite{kklmmt,Firouzjahi:2005dh,Copeland:2003bj,Shandera:2006ax}.} we
have in this case $G\mu_F=8.25\times 10^{-6}$ for F-strings and
$G\mu_D=8.25\times 10^{-5}$ for D-strings, also beyond the
observational bound by one and two orders of magnitude,
respectively. Hence, we see that, though the noncommutative brane
inflation model can successfully realize a large running of the
spectral index and can effectively eliminate the fine-tuning problem
of the parameter $\beta$, the tension of cosmic strings predicted by
this model is not reasonable. In addition, let us consider the
cosmic string tension $G\mu_{cs}$ together with the tensor/scalar
ratio $r$ in this model. In the noncommutative brane inflation
model, also in the case of $N=54$, $\beta=0.225$ and $\mu=0.43$, we
have $r\simeq 1.1$, implying a rather big gravity wave magnitude. We
show the correlation between $\lg (G\mu_{cs})$ and $r$ in the panel
(b) of figure \ref{fig:correlation}, where shaded region come from
varying the parameters $N$, $\beta$ and $\mu$. It can be seen from
this figure that the bound of $r$ from WMAP3 ($r\leq 1.5$ at 95$\%$
confidence level) can be easily saturated by this model.

\section{Fitting to WMAP three-year results}

In this section, we fit the noncommutative D3/$\overline{D}$3-brane
inflation model to the WMAP three year results. For the primordial
density perturbations, if assuming that the primordial fluctuations
are adiabatic with a power law spectrum, the WMAP data require a
spectral index that is significantly less than the
Harrison-Zel'dovich-Peebles scale invariant spectrum, namely a
red-tilted spectrum, $n_s=0.951^{+0.015}_{-0.019}$. On the other
hand, when allowing for a running spectral index, the WMAP data will
favor a blue-tilted spectral index with a large negative running, at
the meantime the data are consistent with large tensor components.
The WMAP results of the running assumption are given by
(\ref{wmap3}), i.e. $n_s=1.21^{+0.13}_{-0.16}$,
$\alpha_s=-0.102^{+0.050}_{-0.043}$, and $r\leqslant 1.5$ at 95$\%$
confidence level. It is remarkable that it is rather hard to produce
a large absolute value of the running spectral index within the
framework of the usual slow-roll inflationary models. Hence, the
confirmation of this suggestive trend is very important for our
understanding of the early universe physics, especially of the
quantum gravity relevant physics. On the other hand, as admitted in
reference \cite{wmap3}, although allowing for a running spectral
index slightly improves the fit to the WMAP data, the improvement in
the fit is not significant enough to require a new parameter.
However, if the running of the spectral index is indeed existent, it
will be an important thing for us to understand the early universe,
so we must attach importance to the running result of WMAP3. Now we
switch to the brane/anti-brane inflationary scenario. For the
original KKLMMT model, there is a fine-tuning problem for $\beta$.
After the announcement of the WMAP three year results, the KKLMMT
model faces very stringent constraints from the WMAP red-tilted
spectrum. It leads to that the allowed range of the parameter space
of the KKLMMT model becomes very narrow, and the fine-tuning of
$\beta$ becomes more severe. For rescuing the
D3/$\overline{D}$3-brane inflation model from the fine-tuning
problem, and for explaining the large running of the spectral index,
the brane inflation model is generalized to the case of
noncommutative space-time. Previous sections have shown that the
noncommutative inflation model can successfully resolve these
problems. Next we shall analyze the model by using the WMAP3
results, and point out the drawbacks of the model.

\begin{figure}[htbp]
\begin{center}
\includegraphics[scale=1.2]{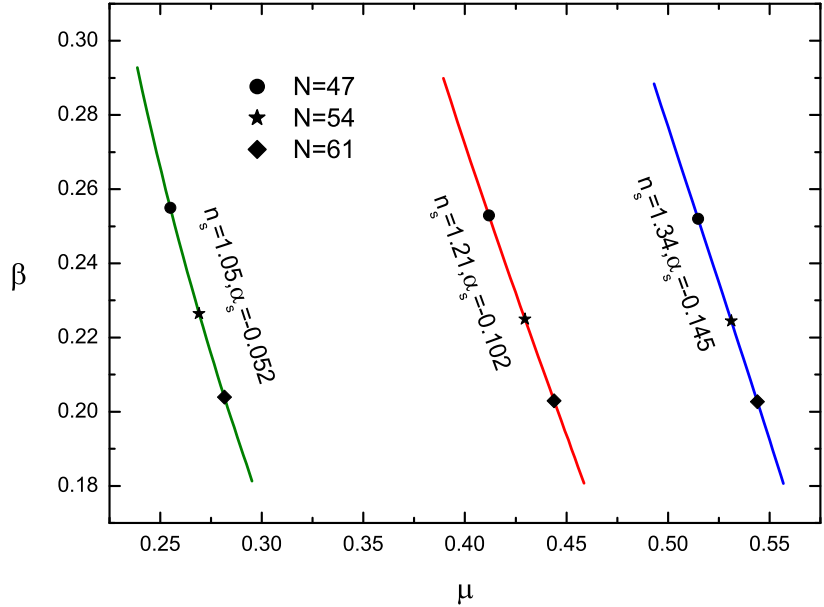}
\caption[]{\small Constraint from the WMAP three-year results. Lines
show the constraint relationship between the mass parameter $\beta$
and the noncommutative parameter $\mu$, in the noncommutative
D3/${\rm \overline{D}3}$ brane inflation, by fitting to the WMAP3
results of the spectral index of the primordial density perturbation
and its running, $n_s=1.21^{+0.13}_{-0.16}$ and
$\alpha_s=-0.102^{+0.050}_{-0.043}$. These lines roughly limit the
range of the parameter space. The range of the $e$-folds before the
end of inflation is from 40 to 70 in this plot. Some locations of
$N$ such as 47, 54 and 61 have been labeled on the
lines.}\label{fig:mubeta}
\end{center}
\end{figure}

\begin{figure}[htbp]
\centering
\begin{center}
$\begin{array}{c@{\hspace{0.2in}}c} \multicolumn{1}{l}{\mbox{}} &
\multicolumn{1}{l}{\mbox{}} \\
\includegraphics[scale=0.9]{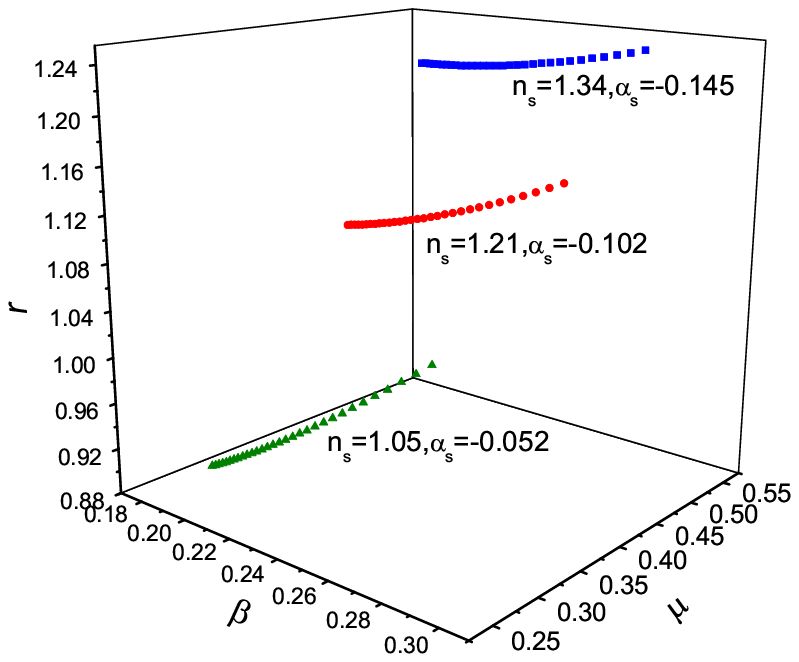} &\includegraphics[scale=0.9]{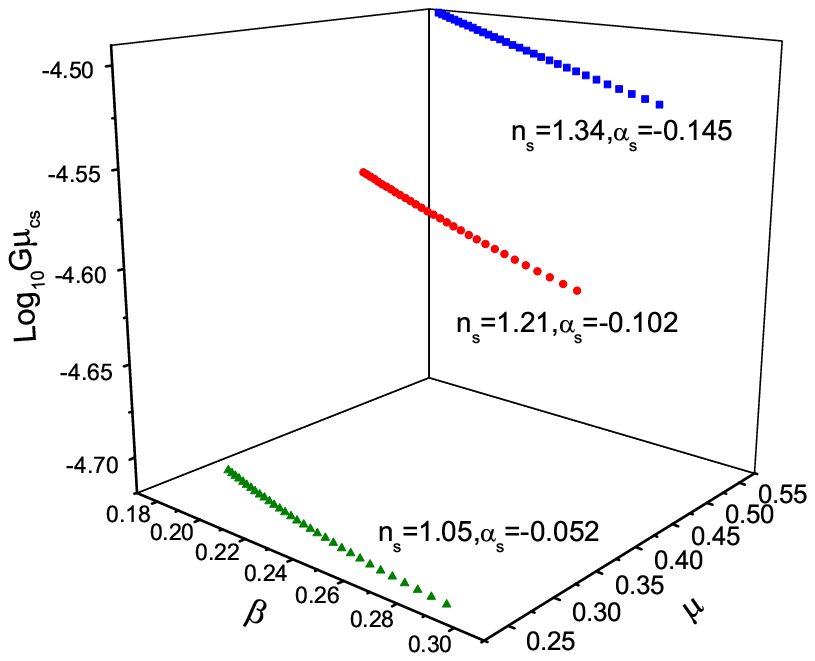} \\
\mbox{\bf (a)} & \mbox{\bf (b)}
\end{array}$
\end{center}
\caption[]{\small Predictions for the scalar-to-tensor ratio $r$ and
the cosmic string tension $\lg (G\mu_{cs})$ of the noncommutative
D3/${\rm \overline{D}3}$ brane inflation model by fitting to the
WMAP3 results $n_s=1.21^{+0.13}_{-0.16}$ and
$\alpha_s=-0.102^{+0.050}_{-0.043}$. Doted-lines correspond to the
central values as well as the 1 standard deviations of the WMAP3
results, respectively. Here we run the number of $e$-folds before
the end of inflation $N$ from 40 to 70. Panel (a) shows the
predictions of $r$. Panel (b) shows the predictions of $\lg
(G\mu_{cs})$.} \label{fig:fitting}
\end{figure}

\begin{table}
\caption{\label{tab:para} Fitting to the WMAP three-year results}
\footnotesize\rm
\begin{tabular*}{\textwidth}{@{}l*{15}{@{\extracolsep{0pt plus12pt}}l}}
\hline\hline
Parameter/Observable&$N=47$&$N=54$&$N=61$\\
\hline
$\beta$&$0.2529^{-0.0009}_{+0.0021}$&$0.2250^{-0.0006}_{+0.0014}$&$0.2030^{-0.0003}_{+0.0010}$\\
$\mu$&$0.411^{+0.104}_{-0.156}$&$0.430^{+0.101}_{-0.161}$&$0.444^{+0.100}_{-0.162}$\\
$\epsilon$&$0.0703^{+0.0066}_{-0.0078}$&$0.0688^{+0.0074}_{-0.0092}$&$0.0678^{+0.0080}_{-0.0103}$\\
$\eta$&$0.0843^{-0.0003}_{+0.0007}$&$0.0750^{-0.0002}_{+0.0005}$&$0.0677^{-0.0001}_{+0.0003}$\\
$\xi^2$&$(7.385^{+0.601}_{-1.200})\times
10^{-12}$&$(3.571^{+0.199}_{-0.478})\times 10^{-12}$&
$(1.864^{+0.066}_{-0.200})\times 10^{-12}$\\
$T_3h^4/M_{\rm Pl}^4$&$(6.619^{+2.972}_{-2.950})\times
10^{-8}$&$(6.807^{+3.113}_{-3.145})\times 10^{-8}$&
$(6.977^{+3.223}_{-3.304})\times 10^{-8}$\\
$\phi_N/M_{\rm Pl}$&$4.449^{+0.219}_{-0.290}$&$4.947^{+0.271}_{-0.372}$&$5.442^{+0.321}_{-0.454}$\\
$r$&$1.13^{+0.10}_{-0.13}$&$1.10^{+0.12}_{-0.15}$&$1.08^{+0.13}_{-0.16}$\\
$G\mu_{cs}$&$(2.57^{+0.52}_{-0.66})\times
10^{-5}$&$(2.60^{+0.54}_{-0.69})\times
10^{-5}$&$(2.63^{+0.56}_{-0.72})\times
10^{-5}$\\
\hline
\end{tabular*}
\end{table}

Using the WMAP three year results for the spectral index and its
running (\ref{wmap3}), we constrain the parameter space of the
noncommutative brane inflation. The main result is shown in figure
\ref{fig:mubeta}. In this figure, we show the allowed region by the
WMAP3 data in the $\beta-\mu$ parameter plane. We run in the plot
the number of $e$-folds $N$ from 40 to 70 for including large
probabilities. For the most likely range of $N$, i.e. $47<N<61$, we
have marked on the plot. According to the fit results, we notice
that the value of $\beta$ has been released to $\sim 1/5-1/4$, which
is a complete relaxation for the fine-tuning; and we find that the
value of $\mu$ is around 0.4, which implies a rather evident
noncommutative effect. In the allowed parameter space, the
noncommutative brane inflation model predicts a rather large
tensor/scalar ratio $r$ and a super large cosmic string tension
$G\mu_{cs}$ as shown in figure \ref{fig:fitting}. As has been
mentioned in the previous section, the predicted cosmic string
tension is beyond the observational bound by two orders of magnitude
when assuming the string coupling $g_s=1$. If we use the fitting
results and the observational bound for cosmic string tension to
determine the string coupling, we have $g_s>10^4$ for D-strings and
$g_s<10^{-4}$ for F-strings. This is rather unnatural. Since the
KKLMMT model is based upon a weak coupled theory (IIB string
theory), the result $g_s\sim 10^4$ is evidently not compatible with
the theory, while the result $g_s\sim 10^{-4}$ seems too weak to be
acceptable.\footnote{If the matter fields live on D3's or
$\overline{D}$3's, then $g_s$ turns out to be exactly $\alpha_{\rm
YM}$. In the special case that $g_s$ is constant, we can identify
$g_s$ with the unified coupling $\alpha_{\rm GUT}$ whose likely
value is around 0.1, so the likely value for $g_s$ should also be
around 0.1 \cite{Polchinski:2004ia}.} This point can be viewed as a
signal of that this model seems somewhat ill-designed.

When specifying the number of $e$-folds, all parameters and physical
quantities of the model can be given according to the WMAP3 data fit
up to one sigma error. We show the fitting results in table
\ref{tab:para}, specifying the number of $e$-folds $N=47$, 54 and
61, respectively. It is interesting to find from the fitting results
in table \ref{tab:para} that the noncommutative brane inflation
model involves some inconsistencies, and we shall analyze these
problems in detail and give some comments.

In table \ref{tab:para}, the slow-roll parameters $\epsilon$, $\eta$
and $\xi^2$ have been determined, i.e. $\epsilon\sim 0.07$,
$\eta\sim 0.08$, and $\xi^2\sim 10^{-12}$. We see that in the
noncommutative model the fit gives $\epsilon\sim\eta$. The values of
D3 brane tension and the inflaton field at the Hubble crossing are
also given, i.e. $T_3h^4/M_{\rm Pl}^4\sim 7\times 10^{-8}$ and
$\phi_N/M_{\rm Pl}\sim 5$. And, the tensor/scalar ratio and the
cosmic string tension predicted by the model are roughly: $r\sim
1.1$ and $G\mu_{cs}\sim 2.6\times 10^{-5}$. It is remarkable that
these features are so different from those of the original KKLMMT
model. For comparison, it is convenient to take an example of the
commutative KKLMMT model. In this case, we take $N=47$ to fit the
spectral index $n_s=0.966$ and get $\beta=6.2\times 10^{-4}$,
$\epsilon=8.3\times 10^{-11}$, $\eta=-0.017$, $\xi^2=3.77\times
10^{-4}$, $T_3h^4/M_{\rm Pl}^4=1.97\times 10^{-17}$, $\phi_N/M_{\rm
Pl}=3.538\times 10^{-3}$, $r=1.33\times 10^{-9}$, and
$G\mu_{cs}=4.43\times 10^{-10}$. Evidently, in the commutative case,
the slow roll parameter $\epsilon$ is very small, so it can be
always negligible and the properties of inflation can be determined
only by the other slow roll parameter $\eta$. Further, we can check
in this case $\beta(\phi_N/M_{\rm Pl})^2=7.8\times 10^{-9}\ll 1$,
thus the approximation used in deriving (\ref{eta}) and (\ref{npp})
works well and the Friedmann equation approximates as $3M_{\rm
Pl}^2H^2\simeq 2T_3h^4$. It can be checked easily that the
approximation will be valid up to about $\beta\sim 1/7$ as shown in
\cite{Firouzjahi:2005dh}. Now, let us check the results in the
noncommutative brane inflation model. In this case, we have
$\epsilon\sim \eta$, so it can not be justified easily whether the
termination condition of inflation can be simply determined by
$\eta\sim -1$. What is more severe is the evaluation of
$\beta(\phi_N/M_{\rm Pl})^2$, that is $\beta(\phi_N/M_{\rm
Pl})^2\sim 0.2\times 5^2\sim 5>1$, so the approximation used in
deriving (\ref{eta}) and (\ref{npp}) is not credible in this case.
In addition, such a big value of $\beta(\phi_N/M_{\rm Pl})^2$ shall
make the dynamical evolution ill-behave. Furthermore, a puzzle
should be noticed. If $\beta(\phi_N/M_{\rm Pl})^2>6$ occurs, the
Friedmann equation breaks down. Unfortunately, this situation often
happens in the fit. For example, taking $N=61$ case, at the central
point we have $\beta=0.2030$ and $\phi_N/M_{\rm Pl}=5.442$, and thus
$\beta(\phi_N/M_{\rm Pl})^2=6.01$; taking $N=54$ case, at the edge
of 1 $\sigma$, we have $\beta=0.2244$ and $\phi_N/M_{\rm Pl}=5.218$,
thus $\beta(\phi_N/M_{\rm Pl})^2=6.11$.

Therefore, although this version of noncommutative brane inflation
can successfully produce a large running of the spectral index and
enormously relax the parameter $\beta$ from the fine-tuning trouble,
one should admit that there are fatal drawbacks in this model. The
main drawback comes from that the incorporation of the brane
inflation scenario and the noncommutative inflation idea is rather
crude. Though it is worth advocating to generalize the brane
inflation scenario to the noncommutative case for resolving the
troubles the KKLMMT model faces, one should find a graceful way to
combine the brane inflation scenario with the noncommutative
inflation framework.


\section{Concluding remarks}

In this paper, we have investigated the noncommutative
D3/$\overline{\rm D}$3 brane inflation scenario in detail. The usual
D3/$\overline{\rm D}$3 brane inflation (KKLMMT) model suffers from
the fine-tuning problem, especially after the announcement of the
WMAP three-year data, the fine-tuning of $\beta$ becomes more severe
(due to the red-tilted spectrum), and only a narrow range of
parameter space for this model is still allowed. On the other hand,
when allowing for a running spectral index, the WMAP three-year data
favor a blue-tilted spectral index with a large negative running.
However, the KKLMMT scenario can not explain the running spectral
index unless accepting a noncommutative generalization. The
noncommutative brane inflation model proposed in reference
\cite{Huang:2006zu} can not only nicely explain the large negative
running of spectral index indicated by WMAP data but also
effectively eliminate the fine-tuning problem in the KKLMMT model.
Though having these advantages, nevertheless, such a noncommutative
KKLMMT model predicts an unacceptable large cosmic string tension
(if assuming $g_s=1$) beyond the observational bound by two orders
of magnitude. A more detailed analysis indicates that there are some
inconsistencies in this model. The most severe problem is that the
value of $\beta(\phi_N/M_{\rm Pl})^2$ derived by fitting to WMAP
three year results is too big to insure the validity of some
approximations in the model and furthermore, the occurrence of
$\beta(\phi_N/M_{\rm Pl})^2>6$ breaks down the Friedmann equation.
This implies that the way of generalizing the brane inflation to the
noncommutative space-time in this model is not delicate enough.
Still, the idea of combining the brane inflation scenario with the
noncommutative inflation framework is interesting and worth
advocating. Hence, it is necessary to find a graceful way of
incorporating these two scenarios and construct a self-consistent
model of noncommutative brane inflation.

\acknowledgments

The author is grateful to Qing-Guo Huang, Miao Li, Jian-Huang She,
Wei Song, Yi Wang and Jingfei Zhang for useful discussions. This
work was supported in part by the Natural Science Foundation of
China.


\end{document}